# Distinct cell shapes determine accurate chemotaxis

Luke Tweedy[1], Börn Meier[2], Jürgen Stephan[2], Doris Heinrich[2,3], Robert G. Endres[1, *]

1. Department of Life Sciences and Centre for Integrative Systems Biology and Bioinformatics, Imperial College, London, United Kingdom.

2. Faculty of Physics and Center for Nanoscience (CeNS), Ludwig-Maximilians-University, Munich, Germany.

3. Leiden Institute of Physics, Leiden University, Leiden, The Netherlands.

* Corresponding Author: Robert G. Endres, r.endres@imperial.ac.uk

**The behaviour of an organism often reflects a strategy for coping with its environment. Such behaviour in higher organisms can often be reduced to a few stereotyped modes of movement due to physiological limitations, but finding such modes in amoeboid cells is more difficult as they lack these constraints. Here, we examine cell shape and movement in starved *Dictyostelium* amoebae during migration toward a chemoattractant in a microfluidic chamber. We show that the incredible variety in amoeboid shape across a population can be reduced to a few modes of variation. Interestingly, cells use distinct modes depending on the applied chemical gradient, with specific cell shapes associated with shallow, difficult-to-sense gradients. Modelling and drug treatment reveals that these behaviours are intrinsically linked with accurate sensing at the physical limit. Since similar behaviours are observed in a diverse range of cell types, we propose that cell shape and behaviour are conserved traits.**

The movement of most animals, from nematode worms [1] to humans [2], can be described by repeated patterns of behaviour in discrete elements, irrespective of underlying details. These stereotyped behaviours are used in response to environmental conditions, e.g. as observed in the hunting behaviour of predatory birds [3]. Similarly on a microscopic scale, cells sense and migrate towards chemoattractants. This process, known as chemotaxis, is remarkably accurate [4, 5], and is fundamental to immune response, wound healing, metastasis, and embryonic development. Despite limited physiological constraints on cell shape, might the actions of eukaryotic cells be described using stereotypes, and might they have a role in chemotaxis?

Traditionally, cell behaviour is described in terms of the underlying molecular processes of the cytoskeleton, adhesion, and signalling. This has met with some success in the study of both fish keratocytes [6] and latrunculin-treated cells [7]. However, keratocytes are not known for efficient chemotaxis and, though latrunculin-treated cells can chemosense, they are immobile. Furthermore, the number of molecular species involved is huge [8], making detailed modelling nearly impossible. An alternative view would be to consider the shape of a cell as an emergent property of all these molecular interactions. Such an approach would have clear advantages; shape is easily accessible experimentally, reduces a cell's complex biochemistry to a single readout, and is more amenable to computational modelling.

In this work, we study cell shapes and their changes in reproducible chemotactic gradients of different steepness. Specifically, we record the shape and movement of over 900 *Dictyostelium discoideum* amoebae, a cell type used for a number of reasons: Firstly, starved cells chemotax accurately towards cyclic adenosine monophosphate (cAMP) for aggregation and subsequent sporulation. Secondly, *Dictyostelium* migration has aroused strong interest recently due to observed "pseudopod splitting", involving branch-like extensions of the cell [9-11]. That these cell shapes are particular to shallow chemical gradients suggests that this mode of behaviour, and the accuracy with which the cell senses, are intimately linked. Finally, their locomotion by pseudopod extension and retraction makes their shape highly irregular. Whilst there have been numerous accounts of pseudopod statistics in the absence of a gradient and in mutants [12-14], so far none have been shape-based. Studies of shape have largely been limited to nuclei [15] and cells with low shape variability [6]. Despite the complexity of cell shape, we are able to account for most of the combined single-cell and phenotypic cell-to-cell variation using only three shape modes. Interestingly, the cells' use of these modes depends on the applied gradient. To gain further insight into the underlying mechanism, we develop biophysical simulations of chemotacting cells, which can quantitatively reproduce behavioural modes in live cells. We use these simulations, along with drug treatments in experiment, to show that cell shape and behaviour are linked with accurate chemotaxis at the fundamental physical limit.

# Results

**Chemotactic index depends on signal-to-noise ratio**

Theory predicts that the fundamental physical limit on the accuracy of chemical gradient sensing is the "perfect absorber" [16]. In this model, ligand molecules are detected on the cell surface and then removed. An absorbing cell is more accurate than a non-absorbing cell by almost an order of magnitude, because ligand molecules are no longer free to unbind and potentially rebind, which adds uncertainty to the cell's measurement. Cells are known to act as absorbers in a number of ways, for example by receptor internalisation [17] and ligand degradation by membrane-bound phosphodiesterase [18]. Both mechanisms are linked to accurate chemotaxis.

The absorber model makes an important prediction for a cell's chemotactic index (CI), defined as the fraction of the total distance travelled in the direction of a chemoattractant. The CI of a cell, a measure of its chemotactic performance independent of migration speed, is predicted to scale with signal-to-noise ratio (SNR), provided by the ratio of the gradient steepness squared over the background concentration. The latter represents the noise arising from the random arrival of ligand molecules at the cell surface by diffusion.

To address this hypothesis, we tracked the movement of many *D. discoideum* cells in a variety of cAMP gradients, produced by a microfluidic device and quantified by addition of an Alexa Fluor red dye with a similar diffusion constant to cAMP [19] (Fig. 1A, B). Note that the fluid carries away with it any cell secretions, thereby removing the impact of cAMP production, as well as degradation by cAMP phosphodiesterase (Péclet number > 1 and [20]). We found that the average CI of the population improved as the SNR increased (Fig. 1C, solid green line). Our results agree closely with previously published data [4] as well as with theory [16] (Fig. 1C, solid black line, using a single parameter fitted to data from [4]). Importantly, neither relative gradient nor concentration alone determine the CI (Fig. 1C, Inset), demonstrating that it is a function of SNR only, and not gradient or concentration individually. This led us to use SNR as our principal environmental parameter in our investigation of cell shape and behaviour, and their role in reaching the fundamental physical limit of gradient sensing.

**Cell shape space is low dimensional**

*D. discoideum* cell shape is extremely variable and has no recurring features that can be used as landmarks when comparing shapes (see Figs. 1A, 2A). We tackle this problem by quantifying shapes using Fourier shape descriptors, which have a number of benefits: they deal with the nonlinear phenomenon of shape naturally, can describe shape in the absence of landmarks, and require only a few Fourier components (Fig. 2B). Additionally, their power spectra are independent of the orientation of cells, bypassing any problems with image alignment (see Supplementary Fig. S2). We then find important modes of variation in the power spectra using principal component analysis (PCA). This technique projects the data into a new set of orthogonal coordinates or principal components (PCs), which are ordered by the size of the variance in the data they account for.

Using this method, we quantified the shapes of 907 cells in environments spanning four orders of magnitude in SNR value. We found that 85% of shape variability can be described using just two shape components and 90% using three (Fig. 2C). These three components correspond approximately to the elongation of the cell (PC 1), pseudopod splitting or bending (PC 2), and cell polarisation (PC 3) (for a test of statistical significance and comparison to an alternative method, see Supplementary Figs. S3 and S4, respectively). The low number of components needed shows that, in spite of the high shape variability relative to other cell types, *D. discoideum* cells occupy a small region in the space of all possible shapes. Furthermore, the shape space has two sharp boundary lines (Fig. S5). The first provides a lower bound on PC 1, and corresponds to a cell with an aspect ratio of one. The second provides a lower bound on PC 2 and corresponds to perfect straightness. These properties strongly confirm our method's ability to discriminate meaningful shape features, independent of orientation and scale.

**Stereotypical cell behaviour**

We observed clear differences in the shapes of cells drawn from low- and high-SNR environments (defined in Fig. S1). Cells at low-SNR had high PC 2 values due to a dominance of split shapes, and low PC 1 values corresponding to small elongation when compared with their respective high-SNR values (Fig. 3A). Such low-SNR cells followed a repeating pattern of shape change, with split pseudopods followed by a period of elongation (Fig. 3B, blue trajectory). In contrast, we observed wedge-shaped, polarised cells predominantly in high-SNR environments, which displayed a remarkable persistency in shape (Fig. 3B, red trajectory). The elongation-then-splitting behaviour in low-SNR cells produces an anticorrelation between PCs 1 and 2 for individual cells over time. This effect, coupled with the high shape-variability of low-SNR cells, leads to a large covariance in PCs 1 and 2 when compared with high-SNR cells (Fig. 3C).

Shape is also a strong indicator of speed. Elongated cells were generally fast-moving when compared with round and split cells (Fig. 3D). Cells slowed as they split pseudopods, and sped up with periods of elongation. Consequently, PC 1 is strongly correlated with speed (Fig. 3E). These characteristic and repeated patterns of shape change and motion represent stereotypical behaviour.

**Simulation reproduces cell shape dynamics**

How do cell shape and behaviour emerge from the underlying biochemistry and biophysics? To investigate the origins of our observed shape modes, we developed computational simulations of a cell boundary evolving in response to the generalised biochemical pathways described by the Meinhardt model [21] (Fig. 4A, B). Each sample point on the simulated membrane is assumed to represent a surface that perfectly absorbs external chemoattractant, and the stimulus of the biochemical pathway is calculated accordingly (see SI for detailed calculations). The force on each point is determined by balancing global membrane tension, local bending tension, cytosolic pressure and the outward force generated by the local concentration of activator. While similar simulations have been used to implement a pseudopod-centred model of chemotaxis [22-24], only our cells absorb ligand molecules in line with recent theory [16]. Whilst the shapes of such simulations differ somewhat from those of live cells (Fig. 4B), they replicate features of the live-cell data. We performed PCA on the shapes of simulated cells, and found that their PCs are similar to those found for live cells (Fig. 4C). Additionally, shape behaviour shows a similar dependence on SNR, with low-SNR cells oscillating between high PC 1 and PC 2 values (Fig. 4D, compare with Fig. 3B).

To elucidate the connection between shape and biochemistry, we compared the autocorrelations of PC 1 from live cells with those of our model (Fig. 5A, for additional correlations see Fig. S6). Both in the data and our model, autocorrelations show sustained oscillation in shape over several periods for low SNR cells, corresponding to changes between elongated and split modes. Oscillations are weaker for high SNR in live cells, and are hardly visible in high SNR simulations. Additionally, the autocorrelation of shape is strikingly similar to that of the local activator in both low and high SNR simulations. Extrema in different shape PCs have unique biochemical profiles, and the protrusion of the membrane broadly follows peaks in activation (Fig. 5B-D). Although our biochemical model species have a number of potential analogues in real cells, a clear candidate for the activator is F-actin, in part due to its fundamental importance in cellular protrusions. Indeed, we observed similar localisations of F-actin in live cells (using LimGFP in [19]) and the activator in simulated cells with similar shapes and conditions (Fig. 5E-G).

**Shape and accuracy are linked**

Using the simulations, we compared the behaviour of normal "wild type" (WT) cells with additional simulated "mutant" cells driven by the same chemistry, but with constrained shapes (see SI for details). As with our live-cell data, the average CI of the WT simulations resembles the curve for the absorber limit (Fig. 6A, solid blue line), recreating an appropriate chemotactic response over the same dynamic range of concentrations as is seen in live cells. In contrast, the CI of the shape-constrained simulations fell short of the absorber limit over a SNR range spanning at least an order of magnitude (Fig. 6A, solid red line).

To validate this prediction in live cells, we considered phospholipase A2 (PLA2) inhibition by p-bromophenacyl bromide (BPB), known to reduce the CI of cells (Fig. 6B, data taken from [25]). To see if this drop in CI could be attributed to shape defects, we treated *Dictyostelium* cells with BPB and projected their shapes into the wild-type shape space. We found that the shapes of BPB-treated shapes occupied only a subsection of the shape-space of wild-type cells, with no extreme values of PC 1 or 2, indicatingan absence of strong elongation or splitting (Fig. 6C). This lower variability in shape is also apparent in the shape-constrained simulations (Fig. 6D). This reduction in CI and shape extrema in both the simulations and data suggests that shape is intrinsically linked to accurate chemotaxis.

# Discussion

We found that *D. discoideum* cells perform chemotaxis using shapes from a low-dimensional shape-space dominated by cell elongation, bending and splitting. Since the use of these shape modes depends on SNR, we propose that stereotypical behaviours are strategies for accurate chemotaxis at the fundamental physical limit. Specifically, the hypothesis of sensing as an absorber allows us to make a number of suggestions for the role of cell shape and behaviour in chemotaxis that discriminate this from other models [26-29].

A potential obstacle in fully reaching the absorber limit is the so-called windshield effect, in which flow of the medium across the cell due to cell movement distorts chemosensation [16]. Although this effect is expected to be small (Fig. S7), pseudopod splitting via PC 2 may allow cells in low-SNR environments to sense sideways, i.e. perpendicular to the direction of movement, thus reducing the windshield effect (Fig. 3A). Following a split, the more favourable pseudopod survives on average [9], leading to a zig-zag movement up the gradient that has been well documented [10]. Second, the observed reduction in speed accompanying oscillations in PC 1 (Fig. 3D, E) may fit a stopping-and-sensing strategy, in which measurements taken by the cell when moving slowly are given more weight in directional decision-making. A related strategy to compensate for the windshield effect would be by secretion of chemoattractant at the rear of the cell, which *Dictyostelium* is known to do using asymmetrically distributed adanylyl cyclase ACA [30].

The link between cell shape and accurate chemotaxis may be even more fundamental. *Dictyostelium* shape mutants are known to exhibit strong chemotactic defects (myoA– mutant in [12], myosin II mutant 3XALA in [31], and AX3:tsuA mutant in [14]). Here, using BPB, a drug known to inhibit chemotaxis [25], we were able to demonstrate a reduction in pseudopod splitting (Fig. 6D). This may impair the above-mentioned strategies, or, alternatively may reduce spatial sampling of the chemical environment. While there is the possibility that the reduced chemotactic index of shape-defect cells (Fig. 6B) is due to signalling and motility impairments, this is very unlikely for two reasons. First, PLA2-inbibition by BPB still allows signalling through the parallel pathway of phosphoinositide 3-kinase (PI3K) [14]. Second, our shape-constrained "mutant" simulations show a similar reduction while having an identical signalling pathway as the wild-type simulations (Fig. 6A). Future work will need to focus on the role of pseudopods and signalling in directional decision-making in chemotaxis.

Amoeboid cell migration based on pseudopods and other irregular cell protrusions is not unique to *D. discoideum*. Such movement has been observed in a variety of cell types including neutrophils [9]) and the spermatozoa of *C. elegans* [32]. Interestingly, the latter achieves motility without actin, emphasising shape and behaviour as fundamentally important. Split pseudopods are also not simply a prerequisite for eukaryotic cell migration: keratocytes with low chemotactic ability move with a single, large lamellipod [6], and at higher speeds than *Dictyostelium*. Hence, we propose that amoeboid shape and behaviour are conserved traits, selected for the evolutionary advantages they yield in chemotaxis. Indeed, the shape space we identified is surprisingly low dimensional (Fig. 2C) despite being relatively unconstrained by the physical properties of the membrane.

Cell shape and its relation to chemotaxis may be clinically relevant, e.g. in monitoring immune activity or detecting metastatic cancer cells. Shape-based methods have demonstrated discriminative power, and have already been used to sort healthy from damaged nuclei [15]. The volume of cell imaging data is expected to grow drastically in the near future, and can be collected in increasingly remote places [33]. The ability to detect disease automatically, using nothing but an image, could be invaluable.

# Materials and Methods

**Experimental Methods.** *D. discoideum* cells of the AX2 strain were starved for 5-7h and then placed in a microfluidic device, as described in [19]. Briefly, the device contained three inputs, left, middle and right, on which the flow could be varied. All inflows contained 17mM K-Na phosphate buffered saline (pH 6.0). The left and right inputs also contained a mixture of cAMP and fluorescein. Gradient profiles ranged between 0-10nM and 0-50nM across 200μm. Video data were sets of differential interference contrast images taken at a resolution of 0.16μm/px every 2s. Cells in which PLA2 was inhibited were treated with 200pM BPB.

**Image pre-processing and shape analysis.** Cell centroid position and outline were recorded over time using a custom-written plug-in for ImageJ. 64 sample points were taken around the outline of each binary image. The coordinates of these points with respect to the centroid were encoded in a complex number such that each shape was a 64 dimensional vector of the form *S = x + iy*. Each shape was transformed into the frequency domain using a fast Fourier transform. Principal component analysis was performed on the power spectra (with the power spectrum *P(f) = |s(f)|$^2$* for the frequency-domain signal *s(f)*) to find the dominant modes of variation (see SI for details and comparison with alternative methods). PCA, Fourier transformation, covariance and correlation calculation and least-squares fits are performed using inbuilt Mathematica functions. Error bars on CI are the standard error for CIs calculated using a range of time windows, to indicate the robustness of the estimates to changes in the amount of data used. Equations describing CI, both with and without the windshield effect, can be found in the supporting information. The autocorrelation of a shape PC for a single cell *APC($\tau$)* is calculated by finding the correlation of the time series of PC values with itself, displaced by $\tau$ time steps. The autocorrelation in local activator *ALA($\tau$)* is found by first calculating the autocorrelation *A($\alpha$,$\tau$)* for each node $\alpha$, counting around the contour from the position at which the activator is at its maximum, displaced in time by $\tau$ time steps. *ALA($\tau$)* is then given by *ALA($\tau$) = (1/N)$\Sigma_N$ $\alpha$ A($\alpha$,$\tau$)*. The period of each autocorrelation is given by *T = 1/f*, where *f* is the native frequency, found by taking the strongest mode of the autocorrelation's Fourier transform.

**Simulations.** Stochastic simulations of cells are discussed in depth in the supporting information. Briefly, we coupled an evolving cell boundary with the three reaction-diffusion equations of the Meinhardt model of dynamic pattern formation [21]. The profile of Meinhardt's activator is directly proportional to the protrusive force acting on each part of the boundary and membrane tension provides a retraction force. The "wild type" and "mutant" simulations are run with identical biochemical parameters, but an increased membrane bending cost constrains the mutant shape. 350 simulations are run for each condition and for a length of time allowing for cells to crawl an average of 19 cell lengths.

# References

1. Stephens, G. J., Bueno de Mesquita, M., Ryu, W. S. & Bialek, W. Emergence of long timescales and stereotyped behaviours in caenorhabditis elegans. *Proc. Natl. Acad. Sci. USA* **108**, 7286–7289 (2001).

2. Hicheur, H., Pham, Q. C., Arechavaleta, G., Laumond, J. P. & Berthoz, A. The formation of trajectories during goal-oriented locomotion in humans. I. A stereotyped behaviour. *Eur. J. Neurosci.* **26**, 2376–2390 (2007).

3. Csermely, D., Bonati, B. & Romani, R. Predatory behaviour of common kestrels (falco tinnunculus) in the wild. *J. Ethol.* **27**, 461–465 (2009).

4. van Haastert, P. J. M. & Postma, M. Biased random walk by stochastic fluctuations of chemoattractant-receptor interactions at the lower limit of detection. *Biophys. J.* **93**, 1787–1796 (2007).

5. Berg, H. C. & Purcell, E. M. Physics of chemoreception. *Biophys. J.* **20**, 93 – 219 (1977).

6. Keren, K. et al. Mechanism of shape determination in motile cells. *Nature* **453**, 475–480 (2008).

7. Parent, C. A., Blacklock, B. J., Froehlich, W. M., Murphy, D. B. & Devreotes, P. N. G protein signalling events are activated at the leading edge of chemotactic cells. *Cell* **95**, 81–91 (1998).

8. Manahan, C. L., Iglesias, P. A., Long, Y. & Devreotes. P. N. Chemoattractant signalling in dictyostelium discoideum. *Annu. Rev. Cell Dev. Biol.* **20**, 223–253 (2004).

9. Andrew, N. & Insall, R. H. Chemotaxis in shallow gradients is mediated independently of ptdins 3-kinase by biased choices between random protrusions. *Nat. Cell Biol.* **9**, 193–200 (2007).

10. Bosgraaf, L. & van Haastert, P. J. M. Navigation of chemotactic cells by parallel signalling to pseudopod persistence and orientation. *PLoS ONE* **4**, e6842, (2009).

11. van Haastert, P. J. M. A stochastic model for chemotaxis based on the ordered extension of pseudopods. *Biophys. J.* **99**, 3345–3354 (2010).

12. Wessels, D., Titus, M. & Soll, D. R. A dictyostelium myosin i plays a crucial role in regulating the frequency of pseudopods formed on the substratum. *Cell Motil. Cytoskel.* **33**, 64–79 (1998).

13. Maeda, Y. T., Inose, J., Matsuo, M. Y., Iwaya, S. & Sano, M. Ordered patterns of cell shape and orientational correlation during spontaneous cell migration. *PLoS ONE* **3**, e3734 (2008).

14. Xiong, Y. et al. Automated characterization of cell shape changes during amoeboid motility by skeletonization. *BMC Syst. Biol.* **4**, 33 (2010).

15. Rohde, G. K., Ribeiro, A. J. S., Dahl, K. N. & Murphy, R. F. Deformation-based nuclear morphometry: capturing nuclear shape variation in hela cells. *Cytom. Part A* **73A**, 341–350 (2008).

16. Endres, R. G. & Wingreen, N. S. Accuracy of direct gradient sensing by single cells. *Proc. Natl. Acad. Sci. USA* **105**, 15749–15754 (2008).

17. Minina, S., Reichman-Fried, M. & Raz, E. Control of receptor internalization, signalling level, and precise arrival at the target in guided cell migration. *Curr. Biol.* **17**, 1164–1172 (2007).

18. Sucgang, R., Weijer, C. J., Siegert, F., Franke, J. & Kessin, R. H. Null mutations of the dictyostelium cyclic nucleotide phosphodiesterase gene block chemotactic cell movement in developing aggregates. *Dev. Biol.* **192**, 181–192 (1997).

19. Meier, B et al. Chemotactic cell trapping in controlled alternating gradient fields. *Proc. Natl. Acad. Sci. USA* **108**, 11417–11422 (2011).

20. Skoge, M. et al. Gradient sensing in defined chemotactic fields. *Integr. Biol.* **2**, 659–668 (2010).

21. Meinhardt, H. Orientation of chemotactic cells and growth cones: models and mechanisms. *J. Cell Sci.* **112**, 2867–2874 (1999).

22. Insall, R. H. Understanding eukaryotic chemotaxis: a pseudopod-centred view. *Nat. Rev. Mol. Cell Biol.* **11**, 453–458 (2010).

23. Neilson, M. P. et al. Chemotaxis: A feedback-based computational model robustly predicts multiple aspects of real cell behaviour. *PLoS Biol.* **9**,

e1000618 (2011).

24. Shi, C., Huang, C-H., Devreotes, P. N. & Iglasias, P. A. Interaction of motility, directional sensing, and polarity modules recreates the behaviours of chemotaxing cells. *PLoS Comput. Biol.* **9,** e1003122 (2013).

25. van Haastert, P. J. M., Keizer-Gunnink, I. & Kortholt, A. Essential role of pi3-kinase and phospholipase a2 in Dictyostelium discoideum chemotaxis. *J. Cell Biol.* **177**, 809–816 (2007).

26. Ueda, M. & Shitaba, T. Stochastic signal processing and transduction in chemotactic response of eukaryotic cells. *Biophys. J.* **93**, 11–20 (2007).

27. Rappel, W. & Levine, H. Receptor noise and directional sensing in eukaryotic chemotaxis. *Phys. Rev. Lett.* **100**, 228101 (2008).

28. Mortimer, D. et al. A bayesian model predicts the response of axons to molecular gradients. *Proc. Natl. Acad. Sci. USA* **106**, 10296–10301 (2009).

29. Amselem, G., Theves, M., Bae, A., Beta, C. & Bodenschatz, E. Control parameter description of eukaryotic chemotaxis. *Phys. Rev. Lett.* **109**, 108103 (2012).

30. Bagorda, A. et al. Real-time measurements of camp production in live dictyostelium cells. *J. Cell Sci.* **122**, 3907–3914 (2009).

31. Stites, J. et al. Phosphorylation of the dictyostelium myosin ii heavy chain is necessary for maintaining cellular polarity and suppressing turning during chemotaxis. *Cell Motil. Cytoskel.* **39**, 31–51 (1998).

32. Nelson, G. A., Roberts, T. M. & Ward, S. Caenorhabditis elegans spermatazoan locomotion: Amoeboid movement with almost no actin. *J. Cell Biol.* **92**, 121–131 (1982).

33. Smith, Z. J. et al. Cell-phone-based platform for biomedical device development and education applications. *PLoS ONE* **6,** e17150 (2011).

## Acknowledgements

We are grateful to Erich Sackmann and Thomas Franosch for their comments on our manuscript. This work was supported by the Biotechnology and Biological Sciences Research Council for funding (L. T.), the Deutsche Forschungs-gemeinschaft (DFG fund HE5958/2-1) and the Volkswagen Foundation grant I/85100 (D.H. and B.M.) and the ERC Starting Grant 280492-PPHPI (R.G.E.).

## Author Contributions

R.G.E. and L.T. conceived the study. D.H. and B.M. designed the experiments. D.H., B.M and J.S. performed the experiments. R.G.E. and L.T. analysed the data and developed the simulations. R.G.E., L.T., B.M. and D.H. wrote the paper.

# Figures

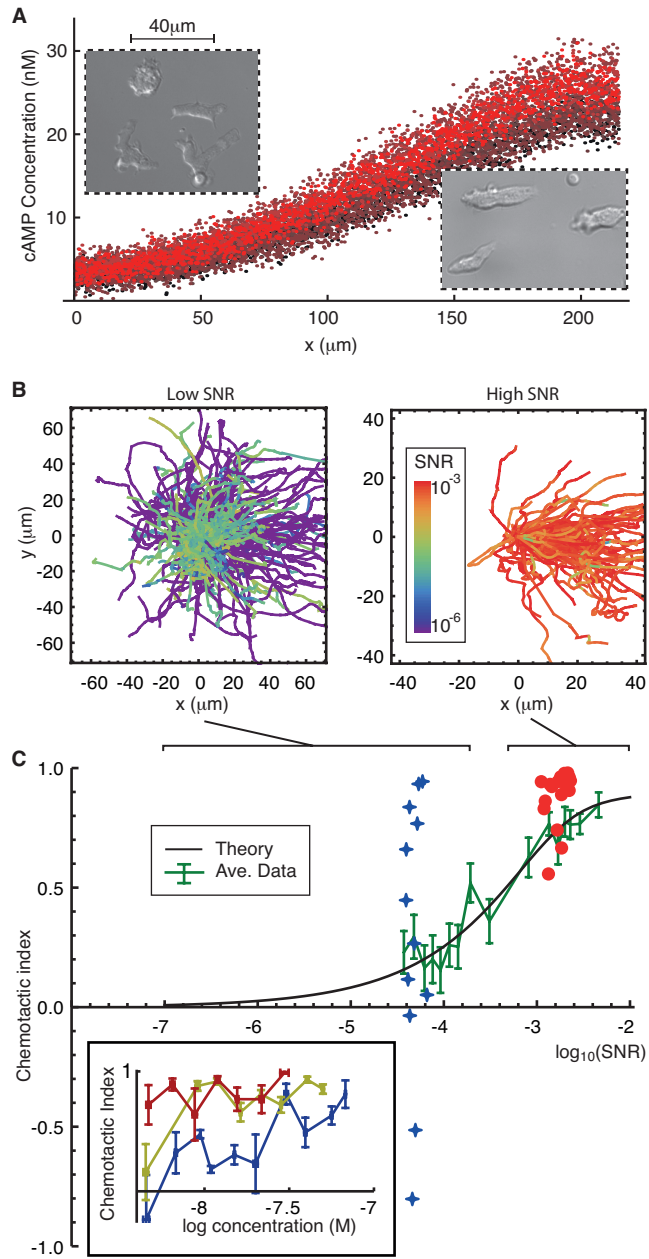

**Figure 1:** Chemotactic index depends on SNR. (A) Gradients of cAMP are produced using a microfluidic device with a flow speed that does not exceed 100μm/s at the depth of the cells. Stable repeats of cAMP concentration profiles are shown in different red shades, with an average standard deviation of about 13% at any given pixel (for experimental details, see supplementary material). (Insets) DIC images of cells exposed to shallow (left) and steep (right) cAMP gradients. Cells in shallow gradients often display multiple or branched pseudopods. (B) Tracks of migrating cells for low- (left) and high- (right) SNR environments, with SNR = $(\nabla c)^2/c_0$, where $\nabla c$ and $c_0$ are gradient and background concentration, respectively (for experimental values, see Fig. S1). Cells bias

their motion in the direction the gradient with increasing SNR. Tracks are coloured based on instantaneous SNR. (C) Chemotactic index is a function of SNR (plotted on a $\log_{10}$ scale). The fundamental physical limit for a static spherical ligand absorber (solid black line) and the mean value of CI are shown (solid green line). Data from two cell trajectories are shown in blue crosses and red circles. To remove non-motile cells, only cells with an average speed above 2.4μm/min were selected. (Inset) CI as a function of concentration for low (blue), medium (yellow) and high (red) relative gradients of 0.0017/μm, 0.010/μm and 0.016/μm, respectively.

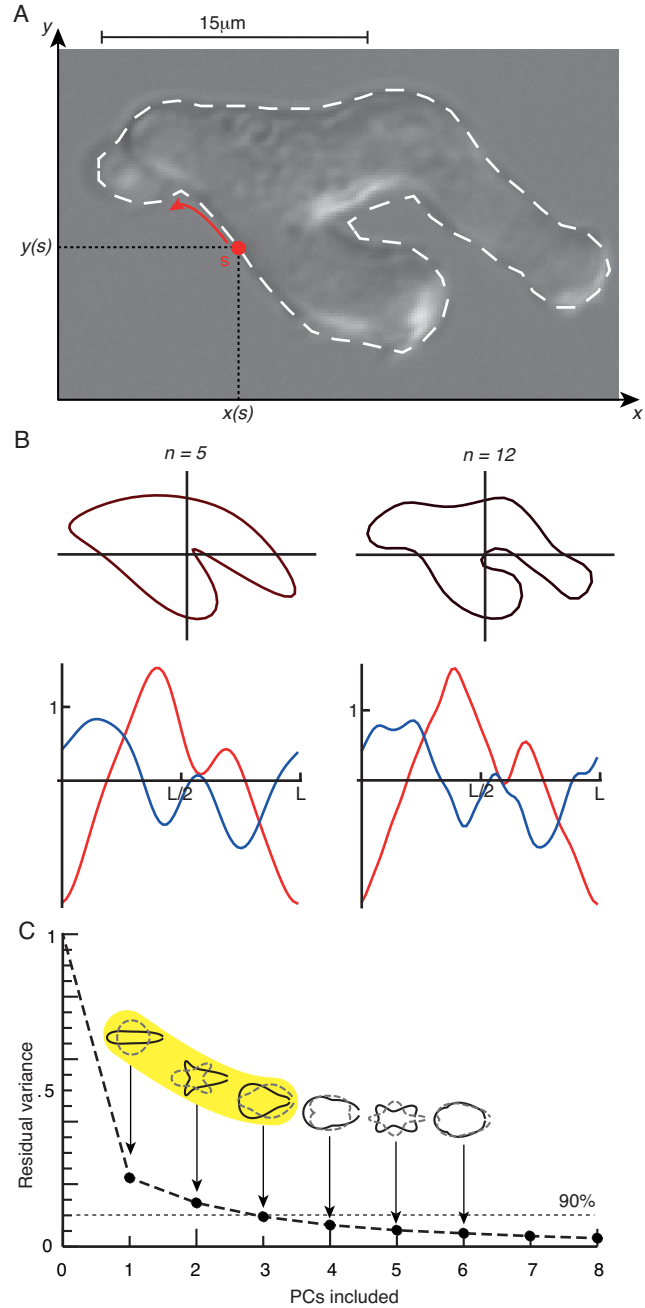

**Figure 2:** Fourier shape descriptors reveal low dimensional shape space. (A) Raw DIC image of a single cell, with the outline taken from the cell superimposed. *x* and *y* co-ordinates are sampled as functions of the distance *s* travelled around the perimeter of length *L*. (B) Reconstructions of the cell outline using *n* Fourier components, with the functions *x,y* shown below. This is the first instance of dimensionality reduction. (C) To find a low-dimensional description of shape, we calculate the power spectra of 3210 cell shapes from 907 cells (see Materials and Methods) and apply principal component analysis (PCA). The residual variance is shown as additional principal components

(PCs) are introduced. Only 3 PCs are required to account for over 90% of shape variability, demonstrating a drastic reduction in dimensionality. These PCs are transformed back into the spatial domain, and are shown as spatial autocorrelations varying around the mean shape with their respective maxima (solid shapes) and minima (dashed shapes). The first three components (highlighted in yellow) describe cell elongation (PC 1), bending or splitting (PC 2), and cell polarisation (PC 3), and are used exclusively from now on.

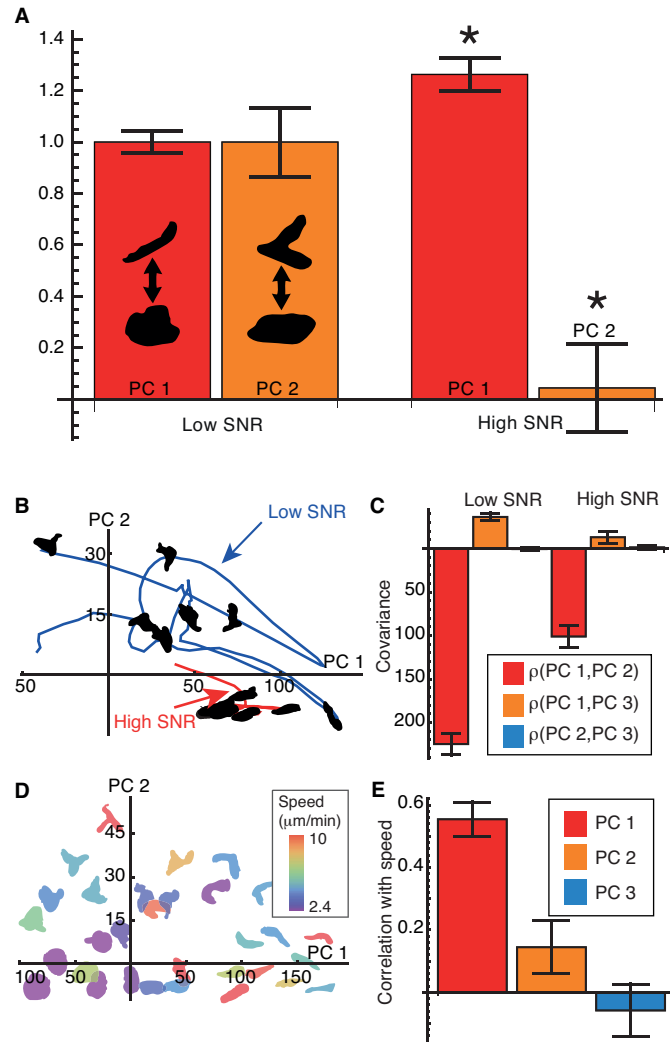

**Figure 3:** Analysis reveals characteristic cell shape and behaviour. (A) Relative differences in cell shape indicate SNR-dependent characteristic shapes. Comparison is based on 967 low- and 361 high-SNR cell shapes. Bars are normalised to the mean of each PC value at low SNR. Absolute values, from left to right, are 44.77, 2.613, 56.32 and 0.114. Error bars show the standard error. Differences in PCs 1 and 2 are statistically significant (*, $p < 0.05$), using a Kolmogorov-Smirnov test and the Bonferroni correction for multiple tests (3 conditions tested). To guarantee independence, only one frame is selected from each cell. Cells are chosen for a migratory speed above an average of 2.4µm/min. (B) Example trajectories projected into shape PCs 1 and 2 for a low- (blue) and a high- (red) SNR cell. The positions of elongated cells at high PC 1 and split cells at high PC 2 confirm our interpretation of the PCs from Fig. 2C. (C) Covariance between pairs of components. In spite of describing independent shape variations, PCs 1 and 2 covary strongly within individual trajectories over time, indicating a stereotypical behaviour. (D) Shape space, coloured according to cell speed, illustrating that cells with low PC 1 and low PC 2 migrate slowly. (E) Correlation of PCs with cell migration speed, demonstrating that high PC 1 values correspond to fast migration. Error bars determined by bootstrapping.

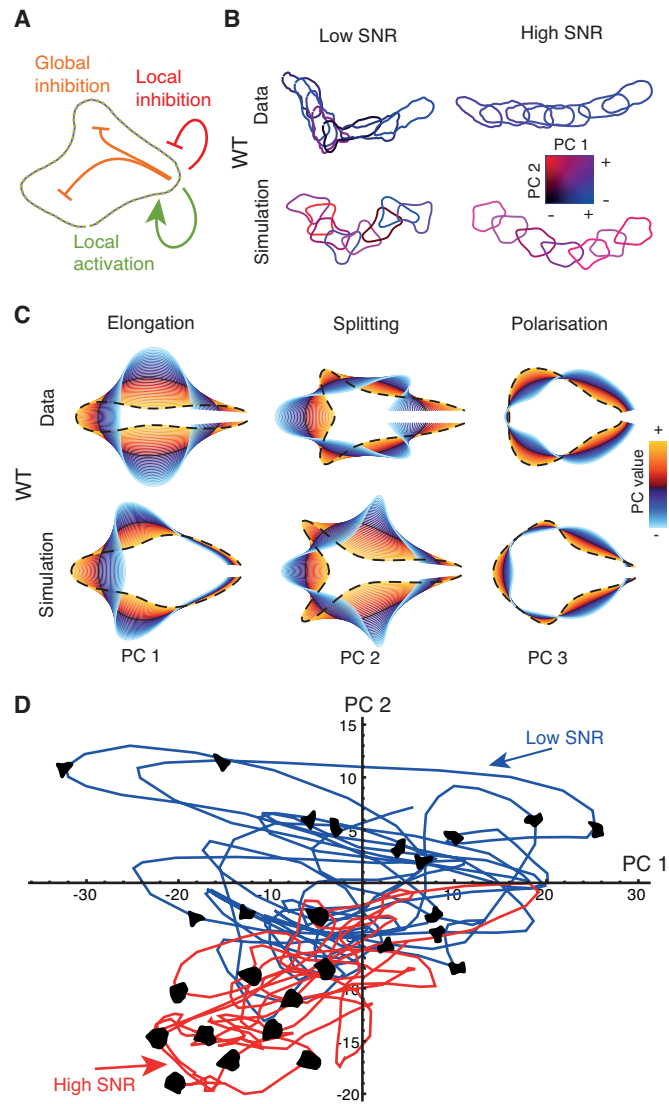

**Figure 4:** Simulation reproduces live cell shape and behaviour. (A) Diagram of the Meinhardt model. A simulated boundary (green nodes) is deformed by forces generated through physical tension and interactions with three simulated species: a locally acting activator that drives protrusion, and both local and global inhibitors, which regulate

activator levels. (B) Example trajectories for wild type (WT) live cells (top) and "wild type" simulations (bottom) for low- (left) and high- (right) SNR. The outlines are coloured according to the value of PCs 1 (red) and 2 (blue). (C) First three PCs of shape for live cells (top) correspond closely to the PCs of simulated cell shapes (bottom). Red to orange contours show increasing addition of a PC to the mean shape. Purple to blue contours show increasing subtraction of a PC. The highest added value of a PC is highlighted (dashed black line). (D) Example trajectories projected into PCs 1 and 2 for low- (blue) and high- (red) SNR simulated trajectories.

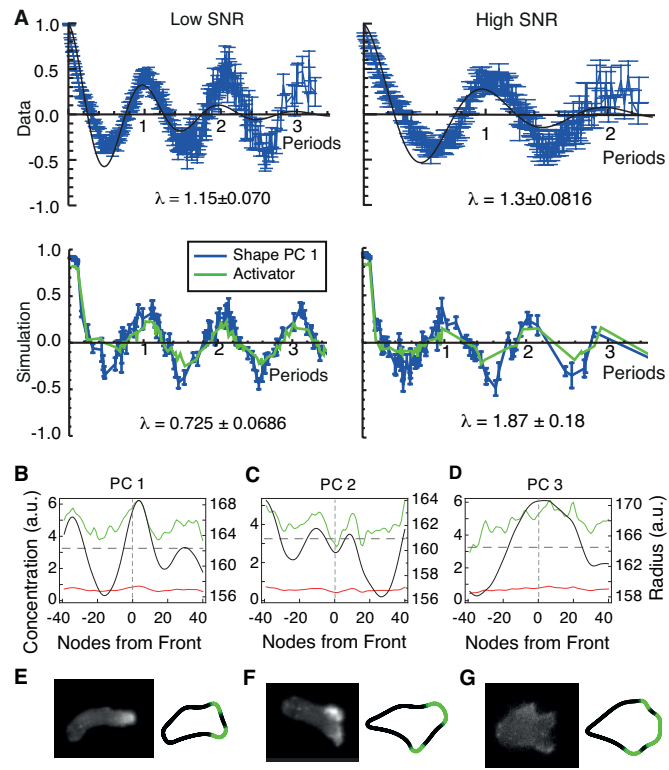

**Figure 5:** Shape gives insight into biochemistry. (A) (Top) Average of autocorrelations in PC 1 for live cells, scaled by native period, for 103 low-SNR cells (left) and 78 high-SNR cells (right). The period of each cell's autocorrelation was obtained by selecting the largest mode in the frequency domain. The median periods for low and high SNR were 173.1s and 239.0s respectively, reflecting the high persistence of cell shape in high-SNR environments. Simulations captured this trend, with median periods of low- and high-SNR simulations of 126.0 time steps and 240.5 time steps, respectively. Error-bars on autocorrelations are large-lag standard errors. (Bottom) Autocorrelations of PC 1 (blue) and of local activator (green) for simulated shallow (left) and steep (right) gradient cells. Individual autocorrelations are aligned according to their fundamental periods, and then averaged. Similar to live cell data, the simulated cells have oscillatory autocorrelations in PC 1, which are less apparent in the steep gradient. For live cell autocorrelations in other PCs, see Fig. S6. (B-D). Concentrations of activator (green), local inhibitor (red) and global inhibitor (dark grey, dashed) and radii (black) are averaged over cells with high values of PC 1 (B), PC 2 (C) and PC 3 (D) and plotted as functions of nodes around the cell contour. Concentration values are shown on the left axis and radii on the right axis. Plots are aligned with the direction of motion in the centre and the rear of the cell at the far left and right. (E-G) Examples of live cells with limGFP-labeled actin (left) compared with simulated cell shapes (right). Cells and simulations in (E) and (F) are exposed to a low SNR. The cell and simulation in (G) are exposed to a high SNR.

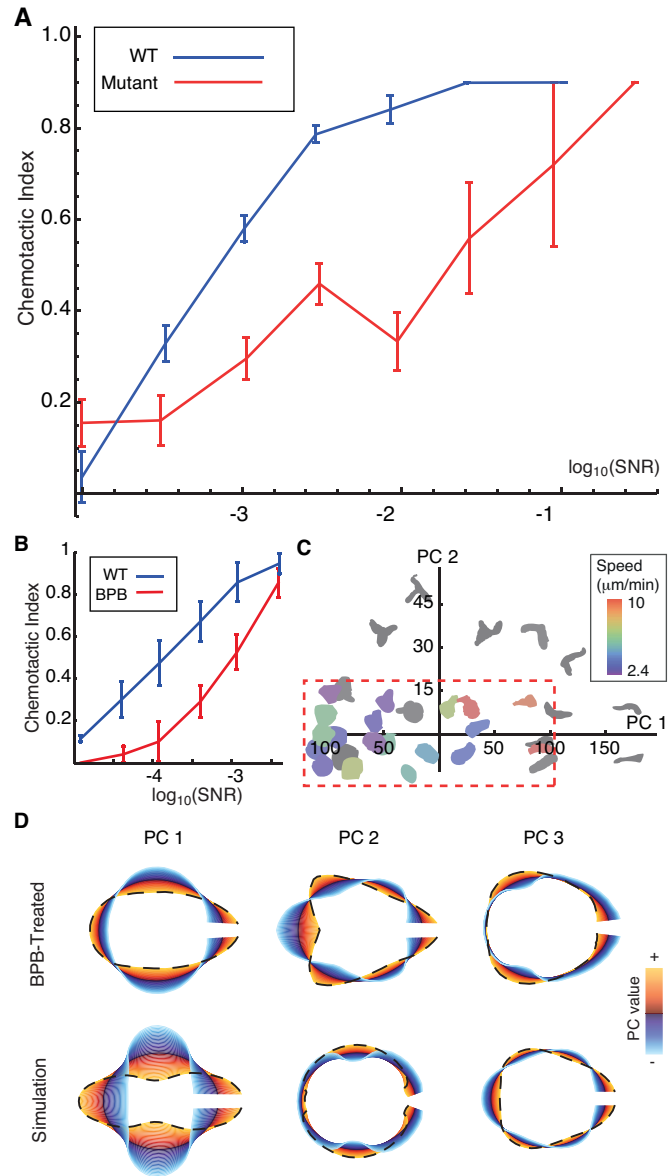

**Figure 6:** Shape is important for accurate chemotaxis. (A) CI as a function of SNR for WT (blue) and mutant (red) simulations. Error bars show the standard error on the mean. (B) The effect of BPB treatment on *Dictyostelium* chemotaxis. Wild type and BPB-treated cells are shown in blue and red, respectively. Graph is based on digitised data from [25], with axis units converted to CI and SNR. (C) The effect of BPB treatment on cell shape. The shapes of BPB-treated cells are shown in colour. Wild type cell shapes are shown in grey for reference. The boundary of the PLA2-inhibited shape space is indicated (dashed red box), and lies inside the shape space boundaries of the wild type data (see Fig. S5). (D) The first three PCs of shape for PLA2-inhibited cells (top) and for shape-constrained simulations (bottom).